\documentclass[bibyear]{aa} 

\newcommand{\new}[1]{{#1}}

\usepackage{graphicx}
\usepackage{natbib}
\usepackage[varg]{txfonts}
\begin{document}

\title{Dark Matter Deprivation in Field Elliptical Galaxy NGC 7507\thanks{Based
    on observations taken at the Gemini Observatory, operated by the
    Association of Universities for Research in Astronomy, Inc., under a
    cooperative agreement with the NSF on behalf of the Gemini partnership:
    the National Science Foundation (United States), the Science and
    Technology Facilities Council (United Kingdom), the National Research
    Council (Canada), CONICYT (Chile), the Australian Research Council
    (Australia), Minist\'erio da Ci\^encia e Tecnologia (Brazil) and SECYT
    (Argentina).}}

\subtitle{}

\author{Richard R. Lane\inst{1}\thanks{rlane@astro-udec.cl}, Ricardo
  Salinas\inst{2} \and Tom Richtler\inst{1}}

\institute{Departamento de Astronom\'ia Universidad de Concepci\'on,
  Casilla 160 C, Concepci\'on, Chile
\and
Department of Physics and Astronomy, Michigan State University, East Lansing,
MI 48824, U.S.A.}

\date{Received.....; accepted.....}

 
  \abstract
  {Previous studies have shown that the kinematics of the field
    elliptical galaxy NGC 7507 do not necessarily require dark matter. This is
    troubling because, in the context of $\Lambda$CDM cosmologies, all
    galaxies should have a large dark matter component.}
   {Our aims are to determine the rotation and velocity dispersion profile out
     to larger radii than previous studies, and, therefore, more accurately
     estimate of the dark matter content of the galaxy.}
   {We use penalised pixel fitting software to extract velocities and
     velocity dispersions from GMOS slit mask spectra. Using Jeans \new{and
       MONDian} modelling we then produce models with the goal of fitting the
     velocity dispersion data.}
   {NGC 7507 has a two component stellar halo, with the outer halo counter
     rotating with respect to the inner halo, with a kinematic boundary at a
     radius of $\sim110''$ ($\sim12.4$\,kpc). The velocity dispersion profile
     exhibits an increase at $\sim70''$ ($\sim7.9$\,kpc), reminiscent of
     several other elliptical galaxies. Our best fit models are those
       under mild anisotropy which include $\sim100$ times less dark matter
       than predicted by $\Lambda$CDM, although mildly anisotropic models
       that are completely dark matter free fit the measured dynamics almost
       equally well. \new{Our MONDian models, both isotropic and anisotropic,
         systematically fail to reproduce the measured velocity dispersions
         at almost all radii.}}
   {The counter rotating outer halo implies a merger remnant, as does the
     increase in velocity dispersion at $\sim70''$. From simulations it seems
     plausible that the merger that caused the increase in velocity dispersion
     was a spiral-spiral merger. Our Jeans models are completely consistent
     with a no dark matter scenario, however, some dark matter can be
       accommodated, although at much lower concentrations that predicted by
       $\Lambda$CDM simulations. This indicates that NGC 7507 may be a
       dark matter free elliptical galaxy. Whether NGC 7507 is completely dark
       matter free or very dark matter poor, this is at odds with predictions
       from current $\Lambda$CDM cosmological simulations. It may be possible
     that the observed velocity dispersions could be reproduced if the galaxy
     is significantly flattened along the line of sight (e.g. due to
       rotation), however, invoking this flattening is problematic.}


   \maketitle
%

\section{Introduction}\label{intro}
The existence of dark matter, or the possibility of a gravitational
interaction differing from that of Newtonian gravity, in spiral galaxies is
not generally disputed. It seems that central cluster galaxies are also
embedded in massive dark haloes \cite[see
  e.g.][]{Kelson02,Richtler08,Schuberth10,Richtler11a}, as are some of the
most massive non-central galaxies in the Fornax and Virgo clusters
\cite[e.g.][]{Napolitano11,Schuberth12}. However, isolated elliptical galaxies
(IEs), and galaxies in less dense environments, are only beginning to be
studied in detail.

Being the isolated counterparts to massive ellipticals in clusters, IEs should
contain information highlighting their evolutionary differences. Globular
cluster (GC) and planetary nebula (PNe) populations are now being employed to
this end, in particular as dynamical probes of galactic haloes
\cite[e.g.][]{Romanowsky03,Douglas07,Napolitano09,Mendez09,Teodorescu10,Richtler11a}.

Techniques for analysing spectra of the stellar halo of galaxies have also
been developed over the last decade or so
\cite[e.g.][]{Kronawitter00,Kelson02,Cappellari04,Samurovic05,Salinas12}, and
more recently in the context of mask spectroscopy
\cite[e.g.][]{Norris08,Proctor09,Foster09,Foster11,Foster13}. Analysis of the
stellar halo light of a galaxy has the advantage that, since the diffuse
stellar halo is being observed, there are no restrictions on image quality, as
is the case for GCs and PNe, while still reaching significantly large
radii. The advantage of mask spectroscopy is that spectra can be obtained at a
variety of position angles within a single observation.

Interestingly, from studies of GCs and PNes, as well as galaxy halo light, it
appears that not all elliptical galaxies require dark matter (DM) to explain
their dynamics and several lower mass IEs only require a minimal dark
halo. This was first investigated at least 20 years ago
\cite[][]{Ciardullo93,Bertin94} and such investigations continue today
\cite[][]{deLorenzi09,Morganti13}. However, numerical simulations based on
Lambda Cold Dark Matter ($\Lambda$CDM), the leading paradigm for Universal
large scale structure formation, show that massive galaxies in all
environments, including IEs, should be dominated by DM
\cite[e.g.][]{Niemi10}. Therefore, if the paucity of DM in some IEs can be
corroborated, and even evidenced in different galactic environments, this
could have serious implications for $\Lambda$CDM cosmological models, at least
in their current form.

One example of an IE that may not require DM to explain its dynamics is NGC
7507, as shown recently by \cite{Salinas12} (hereafter S12). Models with a
cored logarithmic DM halo could also reproduce the observed velocity
dispersion profile, assuming some radial anisotropy, however, these models
only contained a maximum 7.4\% of the total mass as DM within 2 effective
radii ($R_e$). Since the influence of DM in high surface brightness galaxies
can only be unmistakably detected at large radii, measuring the galaxy
kinematics at these large radii is essential. The very interesting result by
S12 has the limitation that the longslit spectra gathered were only able to
probe to radii $R\lesssim1\,R_e$.

In the current paper we extend the work of S12 using GMOS slit masks to cover
a large range of position angles and obtain a velocity field out to
$\sim195''$, $\sim2.6,R_e$ (S12: $R_e\sim75''$). This allows us to search for
rotation signatures of the halo out to large radii (Section \ref{slitvelsec}),
as well as extend the velocity dispersion profile out to $\sim1.7\,R_e$
(Section \ref{veldispsec}). Furthermore, with the velocity dispersion profile
we were able to produce dynamical models out to larger radii than previously
possible, to probe the dark matter content of the galaxy (Section
\ref{modelsec}). Note that our assumed distance to NGC 7507 of
$23.22\pm1.8$\,Mpc (\citealt[][]{Tonry01}, corrected by $-0.16$ mag as per
\citealt{Jensen03}), implies a scale of 112.5\,pc per arcsecond, which we will
use throughout this paper.


\section{Observations and Data Reduction}\label{observationsec}
\begin{figure}
  \centering
  \includegraphics[width=0.485\textwidth]{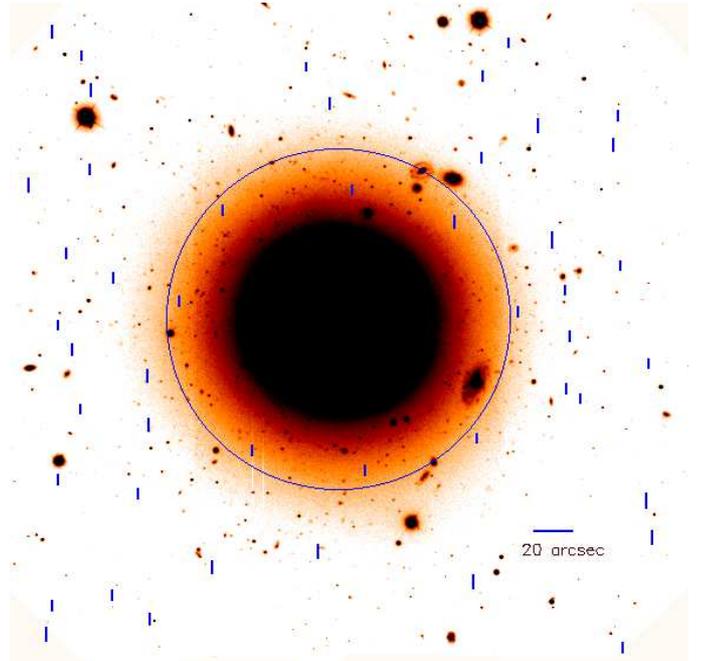}
  \caption{GMOS mask slit positions, centered on NGC 7507, at
    (RA,dec)=($23^{\rm h}12^{\rm m}07^{\rm s},-28^\circ32'23''$). The science
    slit positions and lengths ($4.\!\!''5-8.\!\!''0$), are represented as
    vertical blue lines. All slits are $1''$ in width. The circle indicates
    the largest radius probed by S12. The GMOS field of view is
    $\sim5.\!'5\times5.\!'5$, and a $20''$ bar is shown for scale. North is
    down and East is to the left.}
  \label{maskfig}
\end{figure}

Observations were taken during the night of September 16, 2009, in queue mode,
using the Gemini Multi-Object Spectrograph (GMOS) on the Gemini South
telescope, at Cerro Pach\'on, Chile (Gemini Program GS-2009B-Q-84), during
dark time (4\% Lunar illumination). A single mask with $1''$ slit widths,
centered on NGC 7507, was used with the B600+\_G532 grism, giving a resolution
of $\sim$4.7\AA~FWHM, for consistency with the long-slit study by S12. Slit
lengths ($4.\!\!''5-8.\!\!''0$) and positions were chosen such that no known
objects overlapped any slit, to ensure as pure a galactic stellar population
as possible within the slits. Figure \ref{maskfig} shows the slit positions of
the 47 slits on the slit-mask, overlaid on the GMOS field of view.

Eight exposures of 1700\,s were separated by $2\times1700$\,s sky exposures,
at a blank sky position $13'$ South of the galaxy, and four Cu--Ar arc
exposures, for use in reduction and calibration. This gave us a total exposure
time of 3 hours and 47 minutes, on source. The sky exposures were taken
immediately following the science exposures to ensure changes in sky
illumination were minimised. The maximum time between science and sky
exposures was $\sim1$ minute. Having the blank sky position $13'$ from the
source precluded galaxy light from contaminating our sky spectra. Furthermore,
this also removes the necessity to invoke complex algorithms for removing the
sky background \cite[e.g.][]{Norris08} or rely on slits at large radii which
still contain some galaxy light \cite[e.g.][]{Proctor09}.

The 47 slits were concentrated on the outer parts of the galaxy, where a low
surface brightness, and hence, a low signal is expected. Because of this, the
reduction process should, ideally, minimise the introduction of noise. The 8
science frames were median combined directly as 2D frames using the
IRAF/Gemini routine $gemcombine$. The two sky frames were averaged together,
also as 2D frames, and this ``master'' sky frame was subtracted from the
median science frame. This also removed the bias, without further introduction
of noise. The data from the three GMOS detectors were then mosaiced using the
IRAF/Gemini routine $gmosaic$. The combined and sky subtracted frame was then
wavelength calibrated using the Cu--Ar arc taken after the science exposures,
with the aid of the IRAF task $gswavelength$. The typical RMS of the
dispersion solution was $\sim0.06$\AA. This dispersion solution was then
applied to the sky subtracted science frame using $gstransform$. Extraction of
the spectra was performed with $gsextract$ using all available light in each
slit. All slits were cut to the wavelength range 4800-5500\AA~and rebinned to
the same dispersion with $dispcor$.

\subsection{pPXF}

To extract velocities (Section \ref{slitvelsec}), and velocity dispersions
(Section \ref{veldispsec}), from each slit we employed the Interactive Data
Language (IDL) penalised PiXel Fitting software
\cite[pPXF;][]{Cappellari04,Cappellari12}. The stellar spectral templates we
used were those from the MILES\footnote{http://miles.iac.es/} library
\cite[see also][]{Sanchez06}, chosen because these templates cover the
wavelength range we require, as well as having sufficient resolution for our
purposes. Furthermore, the MILES library was chosen to emulate the study by
S12. For an in-depth discussion of both pPXF and the MILES stellar library,
please see S12.

\section{Slit velocities}\label{slitvelsec}

\begin{figure}
  \centering
  \includegraphics[width=0.45\textwidth,angle=-90]{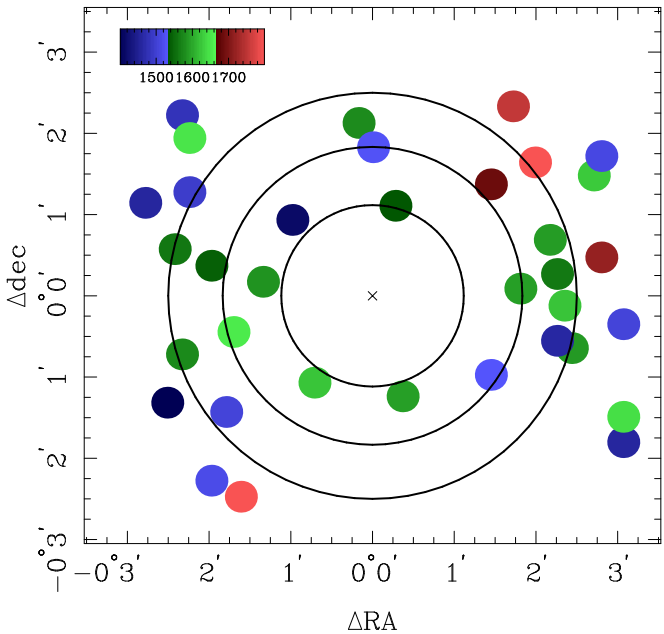}
  \caption{The velocities of the 36 individual slits with good velocity
    estimates. The individual spectra are plotted as large coloured points
    with colours representing their velocities. The legend at the top left is
    given is km\,s$^{-1}$. The orientation has been chosen to correspond to
    that of Figure \ref{maskfig}, so that North is down and East is to the
    left. Large circles delineate the bin edges used in Figure \ref{rotfig} at
    $67''$, $110''$ and $150''$.}
  \label{slitvelfig}
\end{figure}

Line-of-sight radial velocities were estimated for each slit directly using
pPXF. For most slits the signal-to-noise (S/N) was high enough ($\gtrsim5$) to
reliably estimate the velocities (sanity checked with $fxcor$, see Section
\ref{veldispsec}). In a few cases, due to CCD imperfections and because some
slits were too short to provide sufficient signal at a given radius, the
uncertainty in the slit velocity was determined to be too large for the
velocity estimate to be useful. Furthermore, the four slits with the largest
radii, those with $R>195''$, had insufficient S/N to provide realistic
velocity estimates. These slits were removed, and only those with
uncertainties less than $100$\,km\,s$^{-1}$ have been included, leaving 36
slits with reliable velocity estimates. The velocities extracted from each
individual spectrum are represented in Figure \ref{slitvelfig} and listed in
Table \ref{slitveltab}.

\begin{table}
\caption{Velocity estimates for each slit, as measured by pPXF, listed by
  increasing slit radius. Columns are: ID of the slit, radius from the
  galactic centre, line of sight radial velocity and the uncertainty in the
  radial velocity measurement.}
\label{slitveltab}
\centering
\begin{tabular}{c c c c}
\hline\hline
Slit ID & $R$ ($''$) & $v_r$ (km\,s$^{-1}$) & $\delta v_r$ (km\,s$^{-1}$)\\
\hline
157 & 67  & 1539 & 9\\
181 & 72  & 1586 & 11\\
38  & 75  & 1624 & 28\\
177 & 76  & 1419 & 17\\
150 & 78  & 1594 & 18\\
48  & 94  & 1653 & 18\\
62  & 95  & 1594 & 16\\
178 & 96  & 1535 & 38\\
83  & 106 & 1544 & 50\\
168 & 109 & 1529 & 31\\
175 & 111 & 1685 & 20\\
113 & 116 & 1572 & 22\\
67  & 120 & 1564 & 25\\
111 & 121 & 1591 & 75\\
1   & 124 & 1464 & 26\\
138 & 124 & 1621 & 33\\
176 & 127 & 1580 & 18\\
144 & 128 & 1508 & 31\\
33  & 131 & 1577 & 28\\
45  & 132 & 1562 & 23\\
182 & 134 & 1591 & 21\\
15  & 141 & 1496 & 30\\
27  & 143 & 1889 & 65\\
180 & 150 & 1719 & 37\\
19  & 155 & 1257 & 76\\
50  & 162 & 1463 & 27\\
126 & 163 & 1507 & 42\\
120 & 166 & 1650 & 21\\
163 & 166 & 1753 & 52\\
155 & 167 & 1629 & 26\\
78  & 172 & 1844 & 82\\
52  & 172 & 1518 & 58\\
95  & 179 & 1515 & 89\\
103 & 181 & 1485 & 23\\
82  & 185 & 1645 & 34\\
109 & 195 & 1461 & 39\\
\hline
\end{tabular}
\end{table}

\subsection{Rotation}\label{rotsec}

Using long slit spectra, the rotation of NGC 7507 was first reported by
\cite{Franx89}, and later corroborated and extended to larger radii by
S12. The latter measured the rotation of NGC 7507 along the ``minor axis''
(stated as lying on the line with a position angle of $0^\circ$) to have a
peak amplitude of $\sim50$\,km\,s$^{-1}$ at $R\lesssim50''$, dropping to
somewhere between $0-50$\,km\,s$^{-1}$, due to large uncertainties, at
$60''\lesssim R\lesssim80''$.

\begin{figure}
  \centering
  \includegraphics[width=0.53\textwidth,angle=-90]{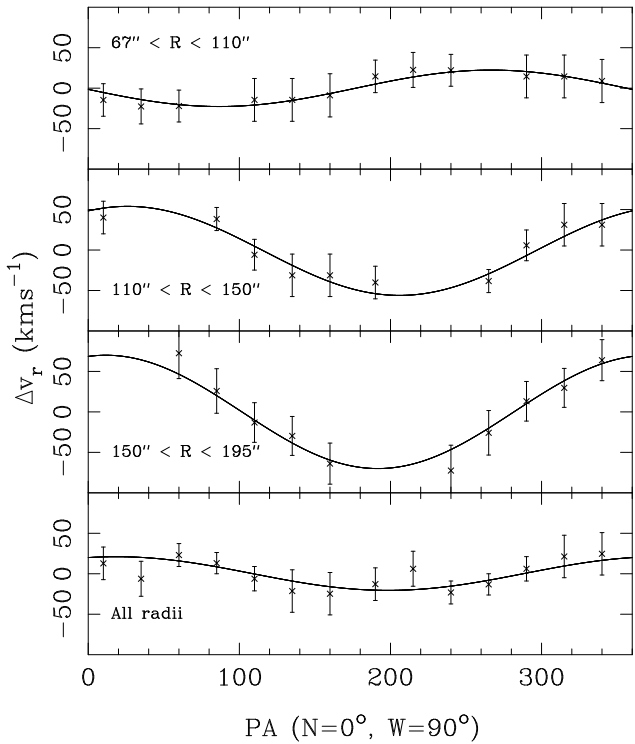}
  \caption{The rotational velocity versus position angle of the halo of NGC
    7507 in three radial bins, shown in each panel, as per the methodology
    described in the text. The error bars represent the RMS errors for each
    PA. The best fit sine curve is overplotted. Note that the $\Delta v_r$
    values calculated by the method described in the text are twice the actual
    rotational velocity. We have taken this into account so that the values
    shown here represent the actual rotational velocity at each PA.}
  \label{rotfig}
\end{figure}

Employing a slit mask, with slits over all position angles (PAs), allows us to
measure the rotation over all PAs, unlike the previous long slit studies. To
measure the rotation, we used a common method \cite[see e.g.][and references
  therein]{Lane09} which involves dividing the slit mask in half at a given PA
and calculating difference between the average velocities in the two
halves. This was performed over PAs in steps of $25^\circ$. Three bins were
used, namely slits with $67''<R<110''$, those with $110''<R<150''$ and those
with $150''<R<195''$. This binning was chosen so that the bins had similar
radial widths and a similar number of slits in each bin. Figure \ref{rotfig}
shows the velocities for each step in PA in each individual bin, as well as
the overall velocities for each step in PA measured with all the slits at all
radii. The best fit sine curve has been overplotted on the resulting rotation
curves. Note that the $25^\circ$ spacing of the steps in PA is arbitrary and
was chosen simply to populate the figure with a sufficiently large number of
data points for visual clarity. Taking larger steps in PA simply reduces the
number of data points and does not affect the results in any way. As can be
seen in Figure \ref{rotfig}, for those PAs where there are no slits the
velocity of those PAs have not been measured.

Serendipitously, the chosen binning also highlights boundaries at which the
rotation of the galactic halo appears to change direction. From Figure
\ref{rotfig}, it is clear that the projected rotational axis of the galactic
halo for radii $67''<R<110''$ lies approximately along the line
PA$=0^{\circ}$, in perfect agreement with the studies by \cite{Franx89} and
S12. While it could be argued that a net rotation of zero may be consistent
with the data, within the uncertainties, the best fit amplitude of rotation
for this inner bin is $\sim22\pm8$\,km\,s$^{-1}$, in agreement with the value
given by S12 at their largest radius of $\sim80''$. For the intermediate bin,
$110''<R<150''$, the projected rotation axis lies approximately along the line
${\rm PA}=120^{\circ}$. It is clear that the direction of rotation at these
intermediate radii {\it has reversed} compared with the inner bin, and has a
rotational velocity of $\sim55\pm7$\,km\,s$^{-1}$. For the outer bin,
$150''<R<195''$, the projected axis of rotation is close to the line ${\rm
  PA}=100^{\circ}$ with the rotation in approximately the same direction as
the the intermediate bin, and a rotational velocity of $\sim70\pm7$. The
overall rotation, measured over all radii, has a projected axis of rotation
lying approximately along the line, ${\rm PA}=80^{\circ}$, similar to that of
the outer halo. At first glance this may seem to indicate that the outer halo
contributes most of the rotational velocity to the galaxy, however, because
the intermediate bin has a very similar rotational axis to the outer bin, and
there are far fewer slits in the inner bin compared to the two outer bins
combined, it is clear that this is only a statistical effect.

To ensure that the binning was not a contributing factor to the rotation
measurements we rebinned the slits at various different radii, making sure
that no bin ever contained fewer than ten slits. Small shifts in the projected
axes and amplitude of rotation were found, however, such shifts are to be
expected, especially for the two inner bins, where the direction of rotation
is reversed. These small differences, and the close agreement at small radii
between our results and those of previous studies, indicate that the binning
is not affecting our results, and, furthermore, that our estimated rotational
amplitudes and axes of rotation are also reliable at large radii. See Section
\ref{discussionsec} for further discussion.

\section{Velocity Dispersion}\label{veldispsec}

\begin{figure}
  \centering
  \hspace{-2mm}
  \includegraphics[width=0.28\textwidth,angle=-90]{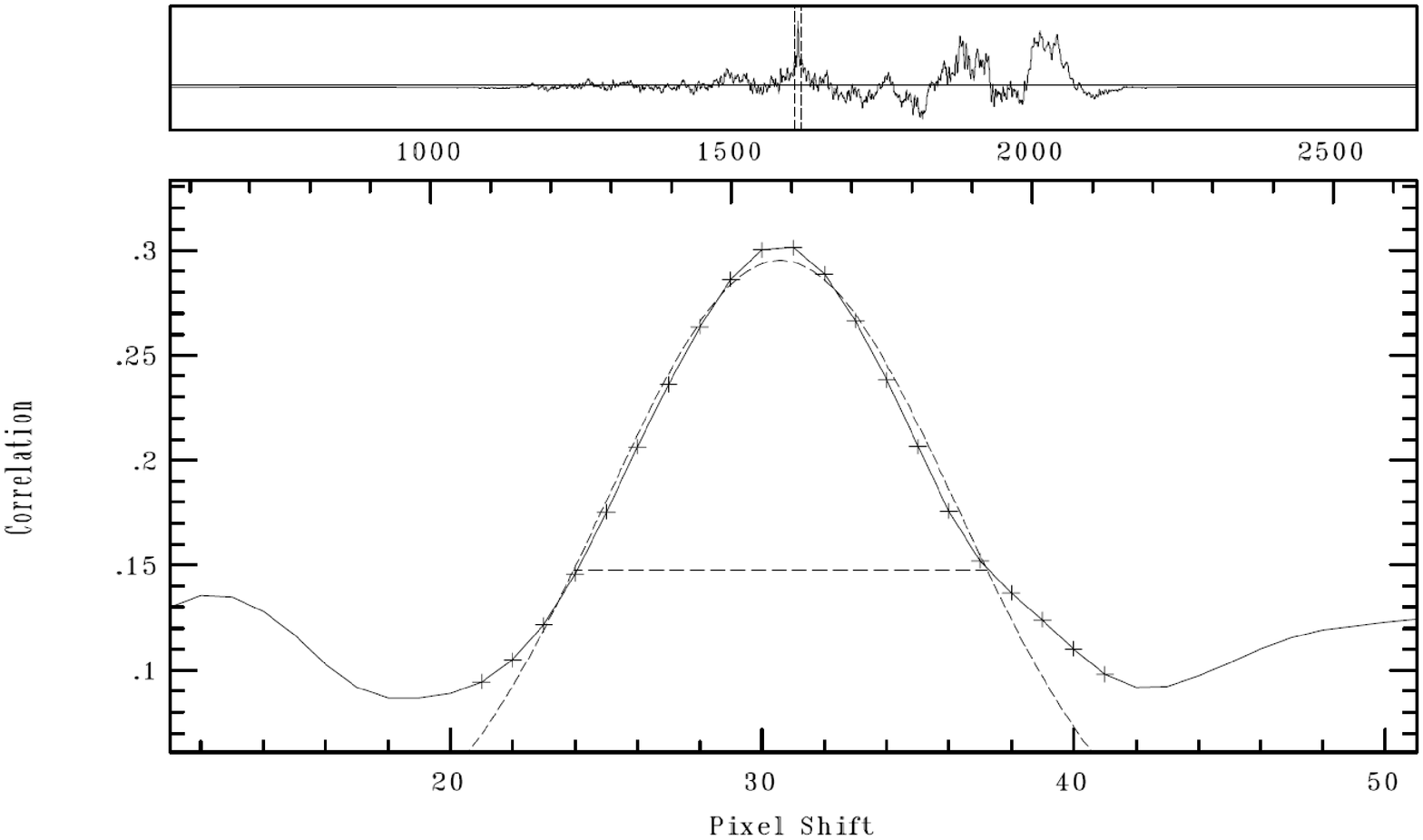}
  \includegraphics[angle=0,width=0.52\textwidth]{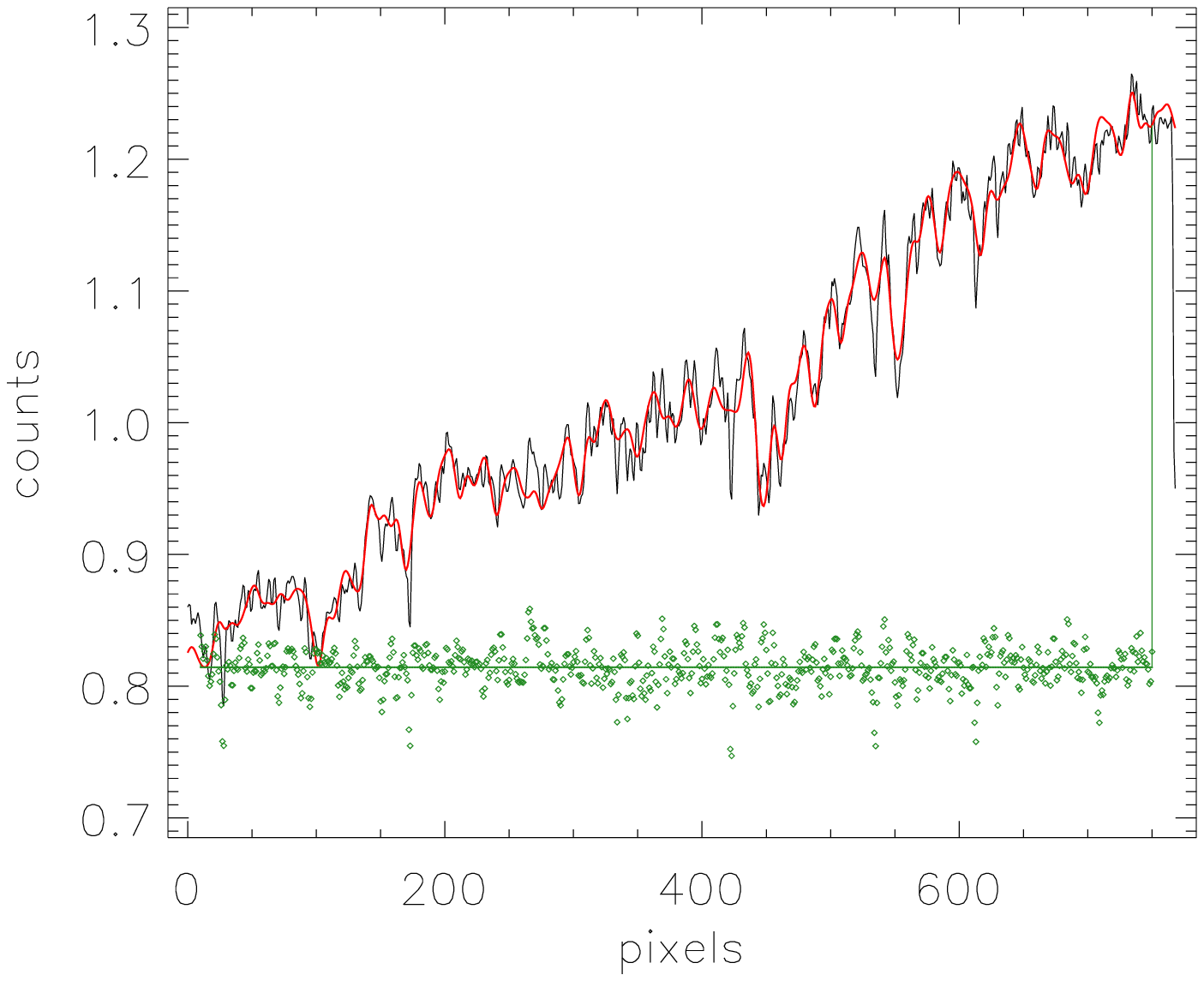}
  \caption{Combined spectrum from slits with IDs 181,38,177 and 150,
    $72''<R<78''$. {\bf Top panel:} Using {\it fxcor} in IRAF, the radial
    velocity for the combined spectrum estimated via cross correlation is
    $1579\pm96$\,km\,s$^{-1}$. {\bf Bottom panel:} The the best fit from pPXF
    (solid red line) to the combined spectrum (solid black line) using 13
    templates from the MILES library, gives a radial velocity of
    $1575\pm6$\,km\,s$^{-1}$. The residuals from the fit are shown as green
    points. Both estimates are consistent with the known values for NGC 7507,
    and are within the uncertainties of each other.}
  \label{testppxffig}
\end{figure}

It was necessary to stack spectra to achieve sufficiently high S/N ($\gtrsim10$)
to allow for accurate velocity dispersion estimates, especially at large radii
where there is less galaxy light in each slit. For the stacking, we shifted
the newly added spectra in steps of 0.1\AA~to locate the shift required to
minimise the velocity dispersion. In this way we ensured that the line widths
in the stacked spectra were not artificially increased due to velocity shifts
in the spectral lines caused by rotation, which would have led to artificial
line broadening, and hence an artificial increase in the velocity
dispersion. To test the veracity of the velocity dispersion minima, we checked
each shifted/stacked spectrum with the IRAF package $fxcor$ to ensure that the
estimated velocity given by $fxcor$ was the same, within the uncertainties, as
that given by pPXF, and also that the minimum velocity dispersion given by
pPXF relates to the minimum FWHM of the cross-correlation function given by
$fxcor$ (see Figure \ref{testppxffig}). The maximum shift for any of the
spectra, as compared to the original spectrum, was 2.5\AA. Note that 2.5\AA~
corresponds to a velocity of $\sim145$\,km\,s$^{-1}$ at 5200\AA, approximately
the central wavelength of our spectra, which is clearly larger than the
rotational velocity. However, the mean shift for all the stacked spectra was
$\sim0.08$\AA, corresponding to a mean velocity of $\sim4.5$\,km\,s$^{-1}$ at
5200\AA, which is compatible with zero, given the uncertainties, as would be
expected under the assumption of solid-body rotation.

To produce the final stacked spectra for analysis we used a ``rolling''
stacking procedure as follows. The first, innermost, spectrum was analysed by
itself, as it had sufficient S/N for an accurate estimate of the velocity
dispersion. Then the spectrum from the slit with the second smallest radius
was analysed to check whether it had sufficient signal. If not, the spectrum
from the slit with the next smallest radius was added (stacked). This process
continued until there was sufficient signal to extract a reliable velocity
dispersion measurement. Once the measurement was performed, the spectrum with
the next smallest radius was added and the spectrum with the smallest radii
from the previously stacked spectrum was removed, this new spectrum was then
checked for sufficient signal, and so on. This procedure produces stacked
spectra that are binned by increasing radius, with varying bin widths, and
with overlapping bin edges, unlike more traditional binning. This technique
was necessary as the available number of spectra was small. Producing binned
spectra in a more traditional way, where spectra are binned without
overlapping bin edges, would have produced velocity dispersion estimates at
only a few radii and removed information from our analysis. Using rolling bins
is a common way to overcome this problem
\cite[e.g.][]{Blom12,Gerke13,Foster13}. It should be noted that the S/N in the
innermost few slits was low, due to short slit lengths, and it was difficult
to extract sufficient signal at these radii, except in the innermost slit. It
was, therefore, useful to include the first slit, which had much higher S/N,
to the innermost four stacked spectra. To stack the spectra we average
combined them using the $scombine$ task within IRAF. At a radius of
$\sim134''$ the uncertainties in the velocity dispersions, as well as the
scatter between individual measurements, began to overwhelm the measurements
themselves. Therefore, we decided to restrict our velocity dispersion
estimates to $R\lesssim134''$. The velocity dispersion estimates are shown in
Table \ref{veldisptab}.

\begin{table}
\caption{Velocity dispersions produced using the rolling stacking technique,
  as described in the text, represented as the open squares in Figure
  \ref{fig:newmodels}. The columns are: 1 - the radii, in arcseconds, of the
  slits combined to produce the final stacked spectra, 2 - the velocity
  dispersion of the final stacked spectra, and 3 - the uncertainty on the
  velocity dispersion estimate.}
\label{veldisptab}
\centering
\begin{tabular}{c c c}
\hline\hline
$R$ ($''$) & $\sigma_v$ (km\,s$^{-1}$) & $\delta\sigma_v$ (km\,s$^{-1}$)\\
\hline
67 & 121.6 & 14.1\\
67-72 & 135.4 & 11.1\\
67-75 & 154.5 & 11.4\\
67-76 & 163.7 & 10.8\\
67-78 & 164.0 & 12.0\\
72-79 & 147.0 & 8.8\\
75-94 & 134.4 & 10.6\\
76-95 & 129.9 & 10.4\\
78-96 & 141.2 & 12.4\\
78-104 & 122.0 & 13.0\\
78-106 & 113.7 & 13.8\\
79-109 & 117.2 & 14.4\\
79-111 & 119.1 & 13.9\\
79-116 & 95.8 & 14.8\\
79-120 & 103.9 & 13.5\\
94-121 & 115.4 & 13.1\\
95-124 & 114.5 & 14.0\\
96-124 & 91.1 & 19.7\\
104-127 & 91.2 & 20.3\\
106-128 & 96.3 & 22.0\\
109-129 & 81.7 & 23.0\\
111-131 & 85.1 & 22.0\\
116-132 & 106.7 & 21.0\\
120-133 & 111.5 & 20.2\\
121-134 & 76.6 & 25.4\\
\hline
\end{tabular}
\end{table}

The velocity dispersion profile is shown in Figure \ref{fig:newmodels}, where
the radii of the individual data points reflect the mean radius of the stacked
slits employed to determine the velocity dispersion. At radii where the data
can be directly compared, our velocity dispersion estimates are completely
consistent with those by S12, who used long slit spectroscopy to determine
velocity dispersions.

Interestingly, it is clear from Figure \ref{fig:newmodels} that there is a
sharp increase in the velocity dispersion at $67''\lesssim R\lesssim70''$. In
the context of a spherically symmetric model, this increase is unphysical. It
has been shown, however, that stellar remnants from mergers can be very long
lived in dwarf spheroidal galaxies (at least up to 10\,Gyr) and, despite being
difficult to observe photometrically, can lead to increased velocity
dispersions \cite[][]{Assmann13a,Assmann13b}. There is a very slight red
colour excess about $10''-15''$ Southwest of the galactic centre
\cite[S12,][]{Caso13}, which may be explained by cold dust \cite[also
  see][]{Temi04}. Apart from this, no obvious signatures of past mergers are
visible in photometric data.

Recent numerical merger simulations by \cite{Schauer14} were performed to try
to explain what they call the ``$\sigma$ bump''. This bump is a flattening of
the line of sight velocity dispersion profile at radii between $\sim1-3\,R_e$
\cite[similar to what is seen, for example, in NGCs 821, 3115 and 4278,
  by:][]{Coccato09,Arnold11,Pota13}. The bump is long-lived -- at the end of
their longest simulation of 9\,Gyr the bump was effectively
unchanged. However, other kinematic remnants, such as shells, dissipate over
much shorter time-scales and are no longer visible after $\sim2$\,Gyr. Note
that the authors simulate mergers between a spiral galaxy and an elliptical
galaxy, and between two spiral galaxies. The bump is more pronounced in the
spiral-spiral mergers, although it is still visible in spiral-elliptical
mergers. It should also be noted that the shape of the bump, at least in the
simulations, does not change with the inclination of the projection, even
though the overall velocity dispersion does decrease with increasing
projection angle. Furthermore, sophisticated $N$-body/hydrodynamic merger
simulations by \cite{Ji14} show features from major mergers last $30\%$ longer
in isolated mergers compared with mergers in cluster environments.

In the simulations by \cite{Schauer14}, this ``bump'' is seen as a velocity
dispersion that is fairly constant with radius. However, in NGCs 821, 3115 and
4278, and now in NGC 7507, there is an increase in the velocity dispersion,
rather than a flattening. In the case of NGC 821 \cite[][]{Coccato09,Pota13},
the increase in velocity dispersion is comparable with what we observe in NGC
7507 ($\sim50$\,km\,s$^{-1}$). Even though the simulations are not a perfect
match to the data in any of these cases, they lend credibility to the idea of
merger activity being the explanation for this deviation from the expectations
of spherical symmetry.

\section{Dynamical Models}\label{modelsec}

\begin{table*}
\caption{Comparison of different best fitting dynamical models including NFW
  halos.}
\label{tab:nfw_models}
\centering
\begin{tabular}{lccccccc}
\hline\hline
$M/L_{\rm stars}$& $\beta$ & $c$ & $\rho_s$ & $r_s$ &$M_{200}^{\mathrm{DM}}$ & $\chi^2$ & $\chi^2$/DOF\\
(M$_{\sun}$/L$_{\sun}$) & & & ($10^{-3}\times$M$_{\sun}$pc$^{-3}$)& (kpc)& (M$_{\sun}$)\\
\hline
\multicolumn{8}{c}{all data}\\ 
\hline 
\noalign{\smallskip}
3.15  &  0*     & -  & -  & -    & -   & 111.43& 2.86\\
3.156 & -0.043  & -  & -  & -    & -   & 111.19& 2.85\\
3.08  &  0*     & 5* & 1.18249*  & 23.5 & $1.85\times10^{11}$ & 100.4 & 2.71\\
3.06  &  0*     & 10*& 6.08990*  & 9.0  & $8.31\times10^{10}$ & 103.1 & 2.79\\
2.96  &  0.3*   & 5* & 1.18249*  & 34.8 & $6.00\times10^{11}$ & 89.6  & 2.39\\
2.91  &  0.3*   & 10*& 6.08990*  & 13.3 & $2.68\times10^{11}$ & 88.1  & 2.38\\
2.78  &  M\L*   & 5* & 1.18249*  & 42.8 & $1.12\times10^{12}$ & 97.3  & 2.56\\
2.78  &  M\L*   & 10 & 6.08990*  & 16.0 & $4.67\times10^{11}$ & 92.8  & 2.44\\
\hline
\multicolumn{8}{c}{no ``bump''}\\            
\hline
3.13 &  0*     & -  & -  & -      & -  &  65.08 & 1.86\\
3.12 &  0.072  & -  & -  & -      & -  &  64.22 & 1.88\\
3.11 &  0*     & 5* & 1.18249* & 12.45 & $2.75\times10^{10}$ & 64.46 & 1.89\\
3.13 &  0*     & 10*& 6.08990* & 1.875 & $7.51\times10^{8}$  & 65.21 & 1.91\\
2.98 &  0.3*   & 5* & 1.18249* & 27.7  & $3.03\times10^{11}$ &  56.5 & 1.66\\
2.94 &  0.3*   & 10*& 6.08990* & 11.0  & $1.52\times10^{11}$ &  55.5 & 1.63\\
2.80 &  M\L*   & 5* & 1.18249* & 36.5  & $6.93\times10^{11}$ &  67.5 & 1.98\\
2.74 &  M\L*   & 10 & 6.08990* & 14.2  & $3.27\times10^{11}$ &  62.6 & 1.84\\
\hline
\end{tabular}
\tablefoot{Quantities with an asterisk mean they have kept fixed during the
  minimisation process. The ``no bump'' panel refers to the quantities derived
  when the velocity dispersion values between 70\arcsec and 80\arcsec are
  removed. M\L~refers to the \citet{Mamon05} anisotropy profile of
  Equation \ref{eq:ml05}.}
\end{table*}

We now use the light model that we produced in S12 to investigate, by simple
dynamical modeling, whether a dark halo is present and what stellar $M/L$
values are required. As in S12, we performed spherical Jeans modeling. Since
NGC 7507 is an E0 \cite[e.g.][]{deSouza04} with a small rotational velocity
(Section \ref{rotsec}), the assumption of sphericity is reasonable, as a first
approximation (see also S12). A compilation of the relevant formulae can be
found in \citet{Mamon05} and \citet{Schuberth10}, and a detailed discussion of
the process can be found in S12 (and references therein).

Our modeling is based on the spherical Jeans equation, which is only valid for
non-rotating systems. Its solutions, however, are considered to be good
approximations for elliptical galaxies, in part because anisotropies in the
inner regions are small, and it has the advantage of simplicity.  Furthermore,
results from more sophisticated dynamical analyses do not differ greatly from
our approach. The bulk of the our uncertainties derive from the uncertainty in
distance and the data quality, rather than the modelling. We multiply the
deprojected light model with a stellar $M/L$ value and calculate line-of-sight
velocity dispersions after adding, or not adding, a dark halo, and with
certain assumptions concerning the anisotropy.

We repeat the analysis of S12, but restricted to a subset of the most
interesting models, without including, for example, the case of logarithmic
halos. Table \ref{tab:nfw_models} gives the results for the selected
models. As in S12, the models which minimise $\chi^2$ are considered as the
best-fitting ones. The inner $3''$ ($\sim330$\,pc) are not considered in the
minimisation given the poor description of the (saturated) light profile in
that range.

\begin{figure*}
  \centering
  \hspace{-0.5cm}
  \vspace{0cm}
  \includegraphics[width=0.98\textwidth]{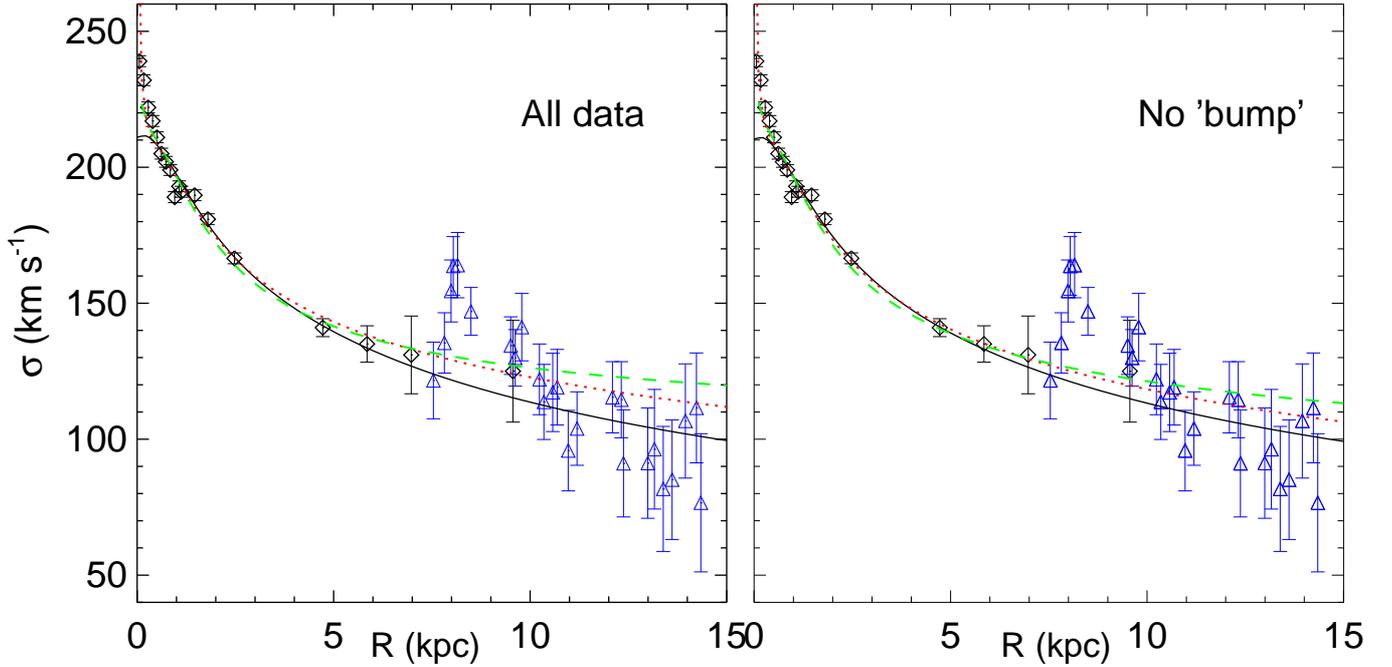}
\caption{Dynamical models. Black open diamonds are the measurements from S12,
  while blue triangles are the velocity dispersions from the current
  work. {\bf Left panel}: Models which consider all the velocity dispersion
  points. The solid black curve is the stars-only model. The red dotted curve
  is the best-fitting model with $\beta=0.3$ and NFW halo with $c=10$. The
  green dashed curve is the model under the anisotropy by \cite{Mamon05}
  allowing the maximum amount of dark mater. {\bf Right panel}: Same as in
  left panel, but excluding the velocity dispersion between $70''<R<80''$ when
  performing the fits.}
  \label{fig:newmodels}
\end{figure*}

The first, and simplest, model is the one where only stars are considered,
under isotropy ($\beta=0$). This model describes the data very well (solid
black curve in the left panel of Figure \ref{fig:newmodels}), implying a
stellar $M/L_R=3.13$. If anisotropy is left as a free parameter, there is a
negligible change from $\beta=0$ to $\beta=-0.043$, without a discernable
improvement to the fit.

The next step is to include a dark matter halo described by a NFW profile
\citep{Navarro97},

\begin{equation}\label{NFWequ}
     \rho(r) = \frac{\rho_s}{(r/r_s)(1+r/r_s)^2},
\end{equation}

\noindent where $\rho_s$ and $r_s$ are the characteristic density and radius,
respectively. As in S12, we restrict the analysis to two concentrations
($c=r_{200}/r_s$) of 5 and 10, both under isotropy and with a mild radial
anisotropy ($\beta=0.3$). All the models including a dark halo provide better
fits (lower $\chi^2$) than the stars-only model, although the amount of dark
matter is very low, and the difference in the goodness of fit is small. Our
most massive halo has a mass of $6\times10^{11}$\,M$_{\sun}$, although a halo
mass of $2.68\times10^{11}$\,M$_{\sun}$ provides a better fit to the data, and
our overall best fit model, which excludes the bump (see below), has a dark
halo mass of only $1.5\times10^{11}$\,M$_\odot$ \new{\cite[which corresponds
    to a dark matter halo concentration, $c_{200}$, of about 6, based on the
    cosmology from WMAP data release 3,][]{Maccio08}.} Even if we assume our
maximum halo mass of $6\times10^{11}$\,M$_{\sun}$, this would imply a stellar
mass of $\lesssim2\times10^{10}$\,M$_\odot$ \cite[e.g.][]{Niemi10}, 10 times
lower than the light profile would suggest (S12). If we adopt a stellar mass
of $\sim2\times10^{11}$\,M$_\odot$ (S12) then this implies a dark halo mass
$\gtrsim10^{13}$\,M$_\odot$, a factor of at least 100 greater than described
by our best fit model. For clarity, Figure \ref{fig:newmodels} (left panel)
only shows the results for the model with $\beta=0.3$ giving the lowest
$\chi^2$. All other models are visually very similar.

As discussed in the previous section, the velocity dispersion profile exhibits
a small increase, or bump, between $70''<R<80''$. We consider now the
influence of this feature in the previous minimisation. The lower panel of
Table \ref{tab:nfw_models} includes the same models described above, but this
time doing the minimisation excluding the bump region. It is clear that the
isotropic NFW models no longer improve the $\chi^2$/DOF when compared to the
dark matter free models, although the mildly radial models continue to give
the lowest $\chi^2$ values.

It is possible that the anisotropy profile of elliptical galaxies is not
constant, but rather increasingly radial at larger radii
\cite[e.g.][]{Dekel05,Hansen06}. For collisionless mergers, \citet{Hansen06}
suggest a universal relation between the anisotropy parameter $\beta$ of the
Jeans equation and the logarithmic slope of the density profile, $\alpha$, of
the form $\beta(\alpha)=1-1.15(1+\alpha/6)$. For our density profile this
means a quick rise up to $\beta=0.4$ at $\sim2$\,kpc, followed by a smooth
increase to $\beta=0.5$ at $\sim15$\,kpc. This behaviour can be well
approximated by the anisotropy profile of \citet{Mamon05},

\begin{equation}\label{eq:ml05}
  \beta_{\rm M\L}(r)=\beta_0\frac{r}{r+r_a},
\end{equation}

\noindent using $\beta_0=0.5$ and $r_a=500$ pc. The dark matter halos which
minimise the $\chi^2$ using this form of anisotropy can also be seen in Table
\ref{tab:nfw_models} for both the complete dataset and the one without the
bump region. As expected from the mass-anisotropy degeneracy, highly radial
orbits allow for higher mass halos, up to $1\times10^{12}$\,M$_{\sun}$ in the
case of $c=5$, although giving systematically poorer fits compared to the
$\beta=0.3$ case. In the case of $c=5$ with the ``no bump'' velocity
dispersions, the fit is even worse than the models without dark matter.

\begin{figure}
  \centering
  \hspace{-0.5cm}
  \vspace{0cm}
  \includegraphics[width=0.49\textwidth]{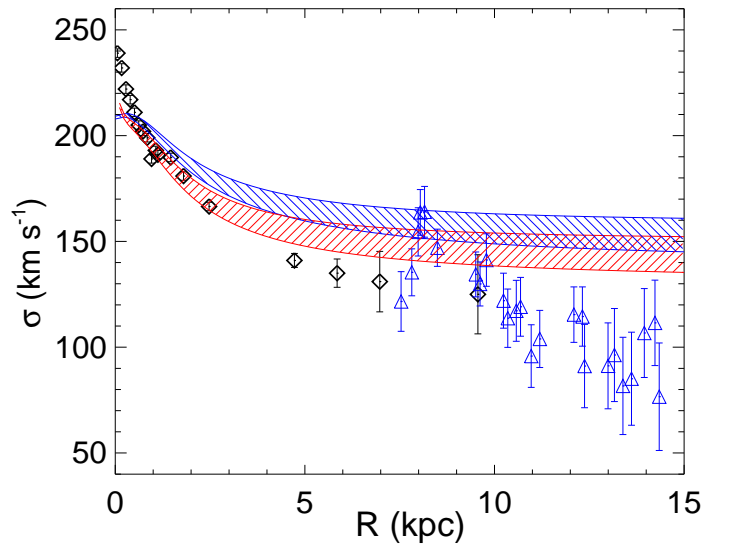}
\caption{MOND models. As in Figure \ref{fig:newmodels}, black diamonds are the
  measurements from S12, while blue triangles are the velocity dispersions
  from the current work. The \new{blue} hatched area covers isotropic models
  with the range of $a_0$ values discussed in the text, while the \new{red}
  area represents the same range for $a_0$ but under the \citet{Mamon05}
  anisotropy profile.}
  \label{fig:mond}
\end{figure}

It is also worth considering whether MOND \citep{Milgrom83} can reproduce
better the observed velocity dispersion profile. In S12, MOND overpredicted
the velocity dispersion at large radii, even when considering the most
favourable cases including a large radial anisotropy. Those results were later
questioned by \citet{Famaey12} (see their Figure 37) who quoted unpublished
work by Sanders, which we cannot assess.

The MOND velocity dispersions are obtained by first calculating the MOND
associated circular velocity from the Newtonian circular velocity using:

\begin{equation}\label{MONDequ}
     v_{\rm M} = \sqrt{v_{\rm N}^2(r)/2 + \sqrt{v_N^4(r)/4 + v_N^2(r) a_0 r}} ,
\end{equation}

\noindent which assumes the simple interpolation formula of \citet{Famaey05}
and where $a_0=1.35^{+0.28}_{-0.42}\times10^{-8}$\,cm\,s$^{-2}$
\cite{Famaey07}. The mass associated to this MONDian circular velocity is then
introduced in the same set of formulae quoted above \cite[see
  also][]{Richtler11b}. Figure \ref{fig:mond} shows the new MOND models
including the extended measurements presented in this paper. As in S12 we
calculate MOND models encompassing the range of accepted $a_0$ values by
\citet{Famaey07}. The hatched blue area indicates the MOND predictions under
isotropy ($M/L_R$=3.0), while the red area uses the anisotropy profile from
Equation \ref{eq:ml05} ($M/L_R$=2.5). Regardless of the assumed anisotropy
profile, or the adopted value of $a_0$, MOND systematically fails to reproduce
the outer velocity dispersion profile.

\begin{figure}
  \centering
  \includegraphics[width=0.5\textwidth]{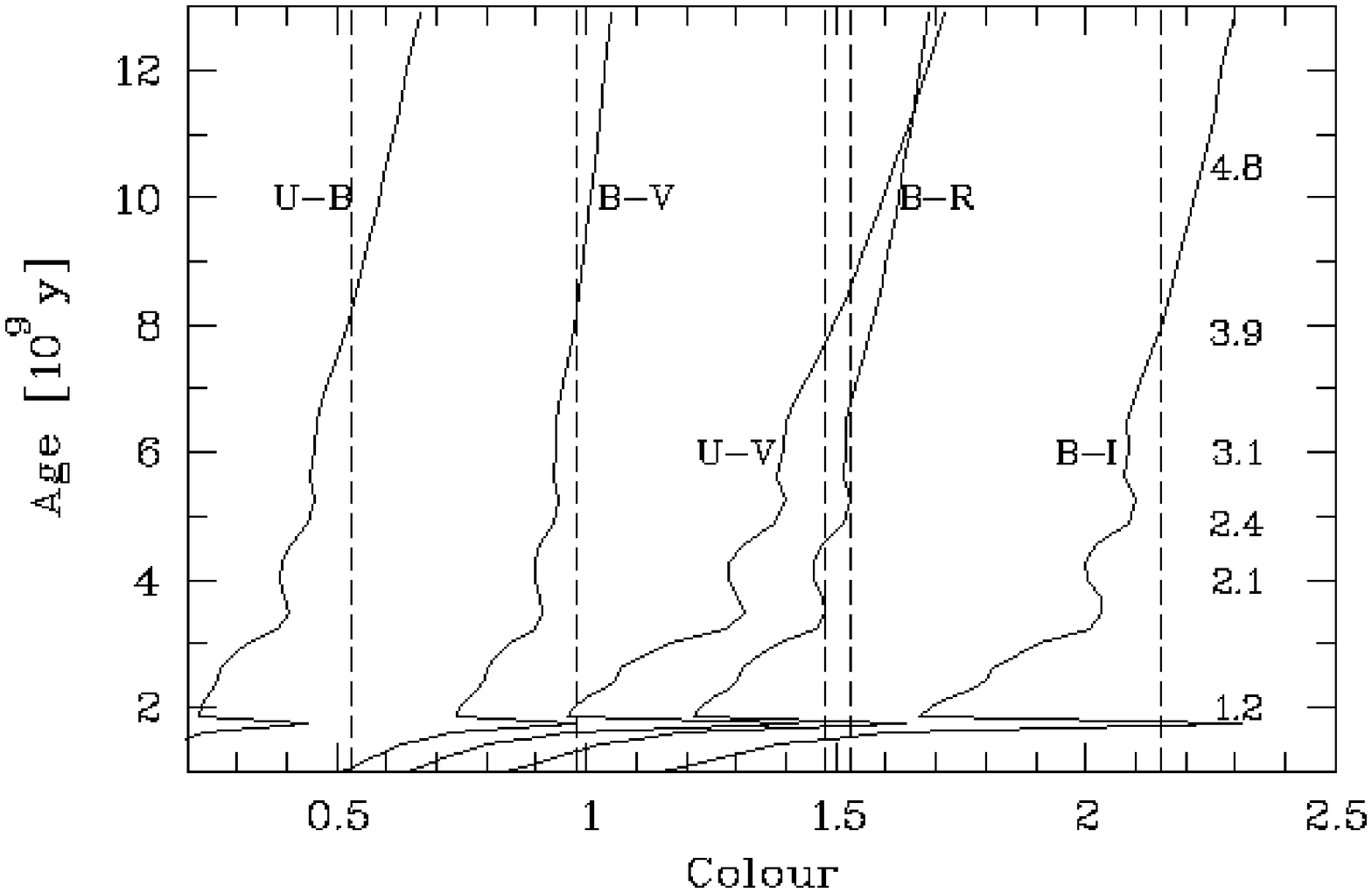}
  \caption{The solid lines represent the models by \cite{Marigo08} and the
    vertical dashed lines are the colours by \cite{Prugniel98} using a $30''$
    aperture. On the right the $M/L_R$ values for a given age
    (\citealt{Chabrier03} IMF -- using M$_{\sun}=+4.42$) are shown. The
    ($B-R$) colour is not really age discriminative, however, an age of
    $\sim8$\,Gyr and $M/L_R\sim4$ is consistent with all other colours ($cf.$
    an age of $\sim4\pm1$\,Gyr by \citealt{Li07}, based on [$B-V$] and [$B-K$]
    colours and the models by \citealt{Bruzal03}). Note that a $M/L_R$ of 3.1
    is well described by our dynamics, but 2.8 is still possible (Table
    \ref{tab:nfw_models}). A $M/L_R$ as small as $\sim2$, to accommodate the
    age estimate by \cite{Li07}, does not match our dynamics.}
  \label{coloursagesfig}
\end{figure}

Another issue to explore are variations in $M/L$. However, there is no
indication from the colour profile that the stellar $M/L$ should vary
(S12). The age of $\sim4\pm1$\,Gyr by \cite{Li07} indicates $M/L_R\sim2.1$
\cite[][]{Bressan12}, however, the colours of the galaxy indicate an age
closer to 8\,Gyr and $M/L_R\sim3.9$ (Figure \ref{coloursagesfig}), and a
$M/L_R$ as low as 2.1 does not match our measured dynamics. Note that in
Figure \ref{coloursagesfig} we have used a \cite{Chabrier03} IMF, which is
fairly bottom-light, in contrast with recent considerations for some
elliptical galaxies \cite[e.g.][]{vanDokkum10,Dutton12,Tortora13}. However,
all such recent studies have focussed on elliptical galaxies in dense
environments. While it seems that massive elliptical galaxies can have more
bottom heavy IMFs in such environments, it is not at all clear that isolated
elliptical galaxies also have bottom heavy IMFs. The age of the galaxy (as
young as $\sim4$\,Gyr, \citealt[][]{Li07}) may accommodate a lower $M/L$
through younger populations. Furthermore, the maximal $M/L$ is that of the
``maximal spheroid'', with no dark matter and isotropy. Any dark matter or
radial anisotropy boosts the central velocity dispersion and requires a lower
$M/L$ as seen in Table \ref{tab:nfw_models}.

As a final note, it is possible that the galaxy is significantly flattened
along the line of sight. Two scenarios may result in this flattening, namely
that NGC 7507 is, in fact, an S0 galaxy seen face on, or that it is rapidly
rotating with the rotational axis close to the plane of the sky (recall that
NGC 7507 is circular along the line of sight). Either of these possibilities
would lower the velocity dispersion at large radii, which may mimic our
measured dynamics. The assumption of flattening is, however, problematic due
to the measured rotation of the galaxy. See Section \ref{discussionsec} for
further discussion.

\section{Discussion and Conclusions}\label{discussionsec}

Using the penalised PiXel Fitting (pPXF) software by \cite{Cappellari04}, we
have analysed 36 GMOS slit mask spectra of the galaxy light of NGC 7507. This
allowed us to calculate a velocity field out to $\sim195''$ ($\sim22$\,kpc),
and a velocity dispersion profile out to $\sim130''$ ($\sim14.5$\,kpc). This
corresponds to between $\sim2.6-6.3\,R_e$ for the velocity field and
$\sim1.7-4.2\,R_e$ for the velocity dispersion profile.

Even though the number of slits we used for measuring the rotation of NGC 7507
is small, our measured rotation is completely consistent with other
studies. We show evidence for the rotational direction changing in the outer
halo, at $R\sim100''$, consistent with results by S12. A change in the
direction of halo rotation at large radii has been observed in other galaxies,
and there are indications that the outer halo of the Milky Way exhibits the
same phenomenon (\citealt{Carollo07}; see also \citealt{Deason11}). It seems
plausible, as suggested by both \cite{Carollo07} and \cite{Deason11} in the
case of the Milky Way, that the counterrotating outer halo of NGC 7507 is the
extant dynamical remnant of an accreted satellite.

While our velocity dispersion estimates agree very well with those by S12, at
radii where our results directly overlap, we have uncovered a peak in the
velocity dispersion at $R\sim70''$, $0.9-2.3R_e$, undetected by those
authors. The most likely explanation for this is also the dynamical remnant of
an accretion event. The slits stacked to estimate the velocity dispersion of
the four bins exhibiting the increased velocity dispersion have PAs
$150^\circ-200^\circ$ apart, i.e. close to opposite sides of the galaxy, as
would be expected by a streaming or bulk motion of stars. Coherent streams can
be difficult to detect photometrically but can lead to increases in velocity
dispersions and can last for at least 10\,Gyr
\cite[][]{Assmann13a,Assmann13b}. Furthermore, simulations of galaxy mergers
by \cite{Schauer14} reveal a ``bump'' in the velocity dispersion between
$1-3R_e$ that can be detected at least 9\,Gyr after the merger event whereas
other merger signatures, such as shells, disappear after $\sim2$\,Gyr. This is
further indicative that the increase in velocity dispersion is due to a merger
remnant and, since no shells or other signatures of a merger are detected,
that the merger occurred more than $\sim2$\,Gyr ago. While this dispersion
increase may have another, as yet unknown, origin, we are working under the
assumption that it is the signature of a merger remnant. We cannot conclude
anything definitive regarding the nature of the interaction. However, from
simulations by \cite{Schauer14} it appears that major mergers impart more of a
bump than minor mergers, as do spiral-spiral interactions, hence, a major
spiral-spiral merger could be the culprit. It is possible, therefore, that
NGCs 821, 3115 and 4278, and 7507 have all formed from spiral-spiral major
mergers. We cannot confirm this possibility and further study is required to
resolve the evolutionary histories of these galaxies.

Using Jeans \new{and MONDian} modelling we have produced various dynamical
models in an attempt to determine whether dark matter can be accommodated
within our data and the age of NGC 7507\new{, or if MOND can reproduce our
  measured dynamics}. Even with anisotropic models we can only accommodate a
small amount dark matter, at least a factor of 100 less than predicted by
$\Lambda$CDM \new{simulations}. \new{These models provide the best fit (lowest
  $\chi^2$/DOF), although the $\chi^2$/DOF of these models is only marginally
  lower than that of our dark matter free models. Furthermore, our MONDian
  models fail to reproduce the measured dynamics at almost all radii.} One
possibility is that the galaxy is significantly flattened along the line of
sight. If NGC 7507 is, in fact, a face-on S0 galaxy, this might also explain
the increase in velocity dispersion at $\sim70''$ since simulations have also
shown that S0 galaxies can be formed from spiral-spiral mergers
\cite[][]{Bekki98}, and, as discussed, spiral-spiral mergers result in a
larger $\sigma$ bump than other merger types. It is, however, notoriously
difficult to distinguish face-on S0 galaxies from elliptical galaxies
\cite[e.g.][]{Tran03,Bamford09,Weinmann09}. Rapid rotation may also be the
result of a major merger. If NGC 7507 is an S0, or is flattened due to
rotation, the change in direction of the inner and outer haloes implies that
the inner and outer regions of the ``disc'' are rotating in different
directions, effectively perpendicular to each other (recall that if it is
flattened it must be flattened along the line of sight and we must, therefore,
be looking at it face-on). We are unaware of any S0 or rotationally flattened
elliptical galaxies where the inner and outer disc regions are rotating with
respect to one another in this manner. We are also unaware of any mechanism
that may cause this kind of dynamical anomaly.

It should be noted that \cite{deLorenzi09} and \cite{Napolitano09} also found
that only very high anisotropy models can accommodate dark matter in the group
elliptical galaxies NGC 3379 and NGC 4494 \cite[see
  also][]{Morganti13}. Adding to the confusion, \cite{Kassin06}, after
decomposing the rotation curves of galaxies into dark and baryonic components,
found that, in the case of the spiral galaxy NGC 157, an NFW dark matter halo
is inconsistent with the shape of the rotation curve, unless baryons are a
significant contributor to the mass within the inner $\sim5$ scale lengths.

It appears, then, that dark matter is not necessarily required to explain the
dynamics of NGC 7507, nor some other elliptical galaxies in various
environments. At the very least the dark halos of these galaxies, including
NGC 7507, are much less massive than expected from $\Lambda$CDM
simulations. Moreover, in at least one case, dark matter is not required to
reproduce the measured rotation curve of a large spiral galaxy. In the context
of $\Lambda$CDM this is an interesting problem that requires a solution. It is
apparent that a systematic study of all galaxies of intermediate luminosities
with apparent paucities of dark matter is needed, and we will be addressing
this in a forthcoming paper.

\begin{acknowledgements}
RRL acknowledges financial support from FONDECYT project No. 3130403. TR
acknowledges financial support from FONDECYT project No. 1100620, and from the
BASAL Centro de Astrofisica y Tecnologias Afines (CATA) PFB-06/2007. TR also
acknowledges a visitorship at ESO/Garching.

We would like to thank Nicola R. Napolitano for his helpful comments and
input, which has lead to a more well rounded paper.
\end{acknowledgements}



\begin{thebibliography}{}

\bibitem[Arnold et al.(2011)]{Arnold11} Arnold, J.~A., Romanowsky, A.~J.,
  Brodie, J.~P., et al.\ 2011, \apjl, 736, L26

\bibitem[Assmann et al.(2013a)]{Assmann13a} Assmann, P., Fellhauer, M.,
  Wilkinson, M.~I., \& Smith, R.\ 2013a, \mnras, 432, 274

\bibitem[Assmann et al.(2013b)]{Assmann13b} Assmann, P., Fellhauer, M.,
  Wilkinson, M.~I., Smith, R., \& Bla{\~n}a, M.\ 2013b, \mnras, 435, 2391

\bibitem[Bamford et al.(2009)]{Bamford09} Bamford, S.~P., Nichol, R.~C.,
  Baldry, I.~K., et al.\ 2009, \mnras, 393, 1324

\bibitem[Bekki(1998)]{Bekki98} Bekki, K.\ 1998, \apjl, 502, L133

\bibitem[Bertin et al.(1994)]{Bertin94} Bertin, G., Bertola, F., Buson, L.~M.,
  et al.\ 1994, \aap, 292, 381

\bibitem[Blom et al.(2012)]{Blom12} Blom, C., Forbes, D.~A., Brodie, J.~P., et
  al.\ 2012, \mnras, 426, 1959

\bibitem[Bressan et al.(2012)]{Bressan12} Bressan, A., Marigo, P., Girardi,
  L., et al.\ 2012, \mnras, 427, 127

\bibitem[Bruzual \& Charlot(2003)]{Bruzal03} Bruzual, G., \& Charlot,
  S.\ 2003, \mnras, 344, 1000

\bibitem[Cappellari \& Emsellem(2004)]{Cappellari04} Cappellari, M., \&
  Emsellem, E.\ 2004, \pasp, 116, 138

\bibitem[Cappellari(2012)]{Cappellari12} Cappellari, M.\ 2012, Astrophysics
  Source Code Library, 10002

\bibitem[Caso et al.(2013)]{Caso13} Caso, J.~P., Richtler, T., Bassino, L.~P.,
  et al.\ 2013, \aap, 555, A56

\bibitem[Carollo et al.(2007)]{Carollo07} Carollo, D., Beers, T.~C., Lee,
  Y.~S., et al.\ 2007, \nat, 450, 1020

\bibitem[Chabrier(2003)]{Chabrier03} Chabrier, G.\ 2003, \pasp, 115, 763

\bibitem[Ciardullo et al.(1993)]{Ciardullo93} Ciardullo, R., Jacoby, G.~H., \&
  Dejonghe, H.~B.\ 1993, \apj, 414, 454

\bibitem[Coccato et al.(2009)]{Coccato09} Coccato, L., Gerhard, O., Arnaboldi,
  M., et al.\ 2009, \mnras, 394, 1249

\bibitem[Deason et al.(2011)]{Deason11} Deason, A.~J., Belokurov, V., \&
  Evans, N.~W.\ 2011, \mnras, 411, 1480

\bibitem[Dekel et al.(2005)]{Dekel05} Dekel, A., Stoehr, F., Mamon.,
  G.~A., et al.\ 2005, \nat, 437, 707

\bibitem[de Lorenzi et al.(2009)]{deLorenzi09} de Lorenzi, F., Gerhard, O.,
  Coccato, L., et al.\ 2009, \mnras, 395, 76

\bibitem[de Souza et al.(2004)]{deSouza04} de Souza, R.~E., Gadotti, D.~A., \&
  dos Anjos, S.\ 2004, \apjs, 153, 411

\bibitem[Douglas et al.(2007)]{Douglas07} Douglas, N.~G., Napolitano, N.~R.,
  Romanowsky, A.~J., et al.\ 2007, \apj, 664, 257

\bibitem[Dutton et al.(2012)]{Dutton12} Dutton, A.~A., Mendel, J.~T., \&
  Simard, L.\ 2012, \mnras, 422, L33

\bibitem[Foster et al.(2009)]{Foster09} Foster, C., Proctor, R.~N., Forbes,
  D.~A., et al.\ 2009, \mnras, 400, 2135

\bibitem[Foster et al.(2011)]{Foster11} Foster, C., Spitler, L.~R.,
  Romanowsky, A.~J., et al.\ 2011, \mnras, 415, 3393

\bibitem[Famaey \& Binney(2005)]{Famaey05} Famaey, B. \& Binney, J. 2005,
  \mnras, 363, 603

\bibitem[Famaey et al.(2007)]{Famaey07} Famaey, B., Gentile, G., Bruneton, J.,
  \& Zhao, H. 2007, \prd, 75, 063002

\bibitem[Famaey \& McGaugh (2012)]{Famaey12} Famaey, B. \& McGaugh, S. 2012,
  Living Reviews in Relativity, 15, 10

\bibitem[Foster et al.(2013)]{Foster13} Foster, C., Arnold, J.~A., Forbes,
  D.~A., et al.\ 2013, \mnras, 435, 3587

\bibitem[Franx et al.(1989)]{Franx89} Franx, M., Illingworth, G., \& Heckman,
  T.\ 1989, \apj, 344, 613

\bibitem[Gerke et al.(2013)]{Gerke13} Gerke, B.~F., Wechsler, R.~H., Behroozi,
  P.~S., et al.\ 2013, \apjs, 208, 1

\bibitem[Hansen et al.(2006)]{Hansen06} Hansen, S.~H., Moore, B., Zemp, M., \&
  Stadel, J.\ 2006, \jcap, 1, 14

\bibitem[Jensen et al.(2003)]{Jensen03} Jensen, J.~B., Tonry, J.~L., Barris,
  B.~J., et al.\ 2003, \apj, 583, 712

\bibitem[Ji et al.(2014)]{Ji14} Ji, I., Peirani, S., \& Yi, S.~K.\ 2014, \aap,
  566, A97

\bibitem[Kassin et al.(2006)]{Kassin06} Kassin, S.~A., de Jong, R.~S., \&
  Weiner, B.~J.\ 2006, \apj, 643, 804

\bibitem[Kelson et al.(2002)]{Kelson02} Kelson, D.~D., Zabludoff, A.~I.,
  Williams, K.~A., et al.\ 2002, \apj, 576, 720

\bibitem[Kronawitter et al.(2000)]{Kronawitter00} Kronawitter, A., Saglia,
  R.~P., Gerhard, O., \& Bender, R.\ 2000, \aaps, 144, 53

\bibitem[Lane et al.(2009)]{Lane09} Lane, R.~R., Kiss, L.~L., Lewis, G.~F., et
  al.\ 2009, \mnras, 400, 917

\bibitem[Li et al.(2007)]{Li07} Li, Z., Han, Z., \& Zhang, F.\ 2007, \aap,
  464, 853

\bibitem[Mamon \& {\L}okas(2005)]{Mamon05} Mamon, G.~A., \& {\L}okas,
  E.~L.\ 2005, \mnras, 363, 705

\bibitem[Macci{\`o} et al.(2008)]{Maccio08} Macci{\`o}, A.~V., Dutton, A.~A.,
  \& van den Bosch, F.~C.\ 2008, \mnras, 391, 1940

\bibitem[Marigo et al.(2008)]{Marigo08} Marigo, P., Girardi, L., Bressan, A.,
  et al.\ 2008, \aap, 482, 883

\bibitem[M{\'e}ndez et al.(2009)]{Mendez09} M{\'e}ndez, R.~H., Teodorescu,
  A.~M., Kudritzki, R.-P., \& Burkert, A.\ 2009, \apj, 691, 228

\bibitem[{{Milgrom}(1983)}]{Milgrom83}
{Milgrom}, M. 1983, \apj, 270, 365

\bibitem[Morganti et al.(2013)]{Morganti13} Morganti, L., Gerhard, O.,
  Coccato, L., Martinez-Valpuesta, I., \& Arnaboldi, M.\ 2013, \mnras, 431,
  3570

\bibitem[Napolitano et al.(2009)]{Napolitano09} Napolitano, N.~R., Romanowsky,
  A.~J., Coccato, L., et al.\ 2009, \mnras, 393, 329

\bibitem[Napolitano et al.(2011)]{Napolitano11} Napolitano, N.~R., Romanowsky,
  A.~J., Capaccioli, M., et al.\ 2011, \mnras, 411, 2035

\bibitem[Navarro et al.(1997)]{Navarro97} Navarro, J.~F., Frenk, C.~S., \&
  White, S.~D.~M.\ 1997, \apj, 490, 493

\bibitem[Niemi et al.(2010)]{Niemi10} Niemi, S.-M., Hein{\"a}m{\"a}ki, P.,
  Nurmi, P., \& Saar, E.\ 2010, \mnras, 405, 477

\bibitem[Norris et al.(2008)]{Norris08} Norris, M.~A., Sharples, R.~M.,
  Bridges, T., et al.\ 2008, \mnras, 385, 40

\bibitem[Pota et al.(2013)]{Pota13} Pota, V., Forbes, D.~A., Romanowsky,
  A.~J., et al.\ 2013, \mnras, 428, 389

\bibitem[Proctor et al.(2009)]{Proctor09} Proctor, R.~N., Forbes, D.~A.,
  Romanowsky, A.~J., et al.\ 2009, \mnras, 398, 91

\bibitem[Prugniel \& Heraudeau(1998)]{Prugniel98} Prugniel, P., \& Heraudeau,
  P.\ 1998, \aaps, 128, 299

\bibitem[Richtler et al.(2008)]{Richtler08} Richtler, T., Schuberth, Y.,
  Hilker, M., et al.\ 2008, \aap, 478, L23

\bibitem[Richtler et al.(2011a)]{Richtler11a} Richtler, T., Salinas, R.,
  Misgeld, I., et al.\ 2011a, \aap, 531, A119

\bibitem[Richtler et al.(2011b)]{Richtler11b} Richtler, T., Famaey, B.,
  Gentile, G., \& Schuberth, Y.\ 2011b, \aap, 531, A100

\bibitem[Romanowsky et al.(2003)]{Romanowsky03} Romanowsky, A.~J., Douglas,
  N.~G., Arnaboldi, M., et al.\ 2003, Science, 301, 1696

\bibitem[Salinas et al.(2012)]{Salinas12} Salinas, R., Richtler, T., Bassino,
  L.~P., Romanowsky, A.~J., \& Schuberth, Y.\ 2012, \aap, 538, A87 [S12]

\bibitem[Samurovi{\'c} \& Danziger(2005)]{Samurovic05} Samurovi{\'c}, S., \&
  Danziger, I.~J.\ 2005, \mnras, 363, 769

\bibitem[S{\'a}nchez-Bl{\'a}zquez et al.(2006)]{Sanchez06}
  S{\'a}nchez-Bl{\'a}zquez, P., Peletier, R.~F., Jim{\'e}nez-Vicente, J., et
  al.\ 2006, \mnras, 371, 703

\bibitem[Schauer et al.(2014)]{Schauer14} Schauer, A.~T.~P., Remus, R.-S.,
  Burkert, A., \& Johansson, P.~H.\ 2014, \apjl, 783, L32


\bibitem[Schuberth et al.(2010)]{Schuberth10} Schuberth, Y., Richtler, T.,
  Hilker, M., et al.\ 2010, \aap, 513, A52

\bibitem[Schuberth et al.(2012)]{Schuberth12} Schuberth, Y., Richtler, T.,
  Hilker, M., et al.\ 2012, \aap, 544, A115

\bibitem[Temi et al.(2004)]{Temi04} Temi, P., Brighenti, F., Mathews, W.~G.,
  \& Bregman, J.~D.\ 2004, \apjs, 151, 237

\bibitem[Teodorescu et al.(2010)]{Teodorescu10} Teodorescu, A.~M., M{\'e}ndez,
  R.~H., Bernardi, F., Riffeser, A., \& Kudritzki, R.~P.\ 2010, \apj, 721, 369

\bibitem[Tonry et al.(2001)]{Tonry01} Tonry, J.~L., Dressler, A., Blakeslee,
  J.~P., et al.\ 2001, \apj, 546, 681

\bibitem[Tortora et al.(2013)]{Tortora13} Tortora, C., Romanowsky, A.~J., \&
  Napolitano, N.~R.\ 2013, \apj, 765, 8

\bibitem[Tran et al.(2003)]{Tran03} Tran, K.-V.~H., Simard, L., Illingworth,
  G., \& Franx, M.\ 2003, \apj, 590, 238

\bibitem[van Dokkum \& Conroy(2010)]{vanDokkum10} van Dokkum, P.~G., \&
  Conroy, C.\ 2010, \nat, 468, 940

\bibitem[Weinmann et al.(2009)]{Weinmann09} Weinmann, S.~M., Kauffmann, G.,
  van den Bosch, F.~C., et al.\ 2009, \mnras, 394, 1213

\end{thebibliography}
\end{document}